\documentclass[prb,floatfix,showkeys,twocolumn,citeautoscript,showpacs,amsmath,amssymb,epsf,aps]{revtex4}
\usepackage{dcolumn}
\usepackage{epsfig}

\usepackage{graphicx}
\usepackage{longtable}
\usepackage{dcolumn}
\usepackage{bm}

\def\BNF{B$_{24}$N$_{24}$ }

\def\SF{$S_{4}$ }
\def\SE{$S_{8}$ }
\def\O{$O$ }

\def\BNR{B$_{96}$N$_{96}$ }

\def\AlNF{Al$_{24}$N$_{24}$ }
\def\AlNE{Al$_{28}$N$_{28}$ }
\def\AlNFE{Al$_{48}$N$_{48}$ }
\def\AlNT{Al$_{32}$N$_{32}$ }
\def\AlNS{Al$_{36}$N$_{36}$ }
\def\AlNR{Al$_{96}$N$_{96}$ }
\def\AlNRA{Al$_{96}$N$_{96}$-I }
\def\AlNRB{Al$_{96}$N$_{96}$-II }

\begin{document}
\preprint{USC/002}

\title{ Electronic structure of fullerene-like cages and finite nanotubes of aluminum nitride}

\author{Rajendra R. Zope$^{1,\dagger}$  and Brett I. Dunlap$^2$}

\affiliation{$^1$Department of Chemistry, George Washington University, Washington DC, 20052, USA}

\affiliation{$^2$ Code 6189, Theoretical Chemistry Section, US Naval Research Laboratory,
Washington, DC 20375}

\altaffiliation{
$\dagger$
Present mailing address : Theoretical Chemistry Section, Naval
Research Laboratory, Washington DC 20375, USA.
}

\date{\today}

\begin{abstract}
   We report density functional study of alternate fullerene-like cage structures and finite 
closed, capped single-wall nanotubes of aluminum nitride.  The cages and nanotubes studied are
modeled as 
Al$_{24}$N$_{24}$, Al$_{28}$N$_{28}$, Al$_{32}$N$_{32}$, Al$_{36}$N$_{36}$,  
Al$_{48}$N$_{48}$, and Al$_{96}$N$_{96}$.  The structure optimization and calculation 
of the electronic structure, vertical ionization
potential, and the electron affinity are performed at the all electron level by 
the analytic Slater-Roothaan method, using  polarized Gaussian basis set of double 
zeta quality.  
All structures are energetically stable with binding energy of about 10-11 eV per AlN pair.
For the larger  Al$_{96}$N$_{96}$, the fullerene like 
cage is energetically less favorable than the {\em two-shell} cluster 
that has  Al$_{24}$N$_{24}$ as an inner shell.
The vertical ionization potential 
and the electronic affinity are in the range 6.7-6.9 eV and 1.5-2.0 eV, respectively.
The binding energy show systematic increase with increase in the length of (4,4) nanotube.
The energy band gap, determined using the $\Delta SCF$ method show that 
these structures are characterized by a fairly large band gap about $4-5$ eV, which is 
however smaller than the gap for the corresponding boron nitride structures.
\end{abstract}

\pacs{ }


\maketitle

\section{Introduction}

   Aluminum nitride (AlN) belongs to the group III-V family of semiconductors and 
possesses number of properties such as low thermal expansion 
(close to that of silicon), high thermal
conductivity, resistance to chemicals and gases normally used in 
semiconductor processing, and 
good dielectric properties, which make it potential candidate for technological applications 
in microelectronics.  Unlike another group III-V semiconductor,
boron nitride (BN), AlN  does not exist in the layered graphitic form, which 
can be rolled up to form nanotubes. Nevertheless,
a few reports of AlN nanotube and nanowire synthesis have appeared in literature in
recent years\cite{Tondare02,Bala04,Expt_JACS,N_Bala04,T04}.
%
 In the earliest report\cite{Tondare02}, the AlN nanotubes (AlN-NT) and nanoparticles
were synthesized by dc-arc plasma method and were characterized by the transmission 
electron microscope (TEM). The nanotubes were measured to be 500-700 nm in length 
with 30-200 nm diameter, and the nanoparticles had diameters in the 5-200 nm size range.
The field emission pattern from tungsten tips coated with AlN  tubes was measured
and was attributed to the tubes having  open ends.  
Using the same dc-arc plasma method,
helical and twisted AlN nanotubes were reported by the same group\cite{Bala04}. The 
nanotubes 
were dispersed on graphite substrate and analyzed by scanning tunneling microscopy. 
The tubes were reported to have an average diameter of 2.2 nm and lateral dimension of 
about 10 nm. The interatomic distance between two Al-Al or N-N atoms was measured to be 
3.2 \AA\, and from I-V curves the AlN-NT were argued to be metallic in character\cite{Bala04}.
In another study\cite{N_Bala04}, same group reported synthesis of AlN nanoparticles in 
the size range 15-18 nm and of nanowires of 500-700 nm in length with diameters in 
the 30-100 nm size range. In a subsequent report, these nanowires  were interpreted to 
be nanotubes\cite{T04}.
 Synthesis of faceted hexagonal AlN nanotubes has also been reported\cite{Expt_JACS}.
 These AlN-NT
were characterized by TEM and are found to be of a few micrometers in 
length and 30-80 nm in diameter. Most tubes were found with both ends open.
There are also a few reports of theoretical calculations\cite{ZZ03,AZaho04,BZaho04}. 
  Zhang and Zhang\cite{ZZ03} considered two model Al$_{27}$N$_{27}$ structures, one  
for  an AlN nanowire
and one for an AlN nanotube at the Hartree-Fock level using the 3-21G\* basis set.
They noted that AlN bond length in finite AlN-NTs, modeled by the Al$_{27}$N$_{27}$ tubule,
alternates between 1.76 \AA\, and 1.77 \AA\, and posses large energy gap of 10.1 eV 
between the highest occupied molecular orbital (HOMO) level and the lowest 
unoccupied molecular orbital (LUMO) level. This rather large gap of 10.1 eV
is  an artifact of the Hartree-Fock approximation which models exchange effects exactly 
but ignores correlations effects.   Our calculations, on the other hand, calculate
the band gap by the so called $\Delta SCF$ where  the first ionization potential 
is subtracted from the first electron affinity. 
In density functional theory for finite systems, this is known to provide better 
approximation to 
the band gap than the significantly underestimated HOMO-LUMO eigenvalue difference.
Another theoretical calculation by Zhao and coworkers\cite{AZaho04,BZaho04}
have used density functional theory in the local density approximation  (LDA)
and generalized gradient approximation (GGA) to study strain energy (energy required 
to curl up nanotube from graphite like planar sheet) and stability 
of selected single-walled AlN-NT. These 
calculations employed a localized numerical orbital basis and psuedopotential 
for description of ion cores.  These authors noted HOMO-LUMO gaps of 
3.67 and 3.63 eV for (5,5) and (9,0) tubes, respectively. Simulations at
elevated temperatures at the level of LDA indicated stability of single
walled AlN-NT at room temperature. 
 In the present paper, we report the study of fullerene analogue of AlN cages and finite $(4,4)$
single-walled AlN nanotubes using analytic density functional theory. Unlike 
AlN-NT, only one study has  so far addressed AlN nitride cage structures. 
%
The study was performed using density functional theory and
indicated  possible existence of  Al$_{12}$N$_{12}$,  Al$_{24}$N$_{24}$, 
and Al$_{60}$N$_{60}$ on the basis of energetics and vibrational stability\cite{AlN_cage2}.
The present study is motivated partly by these works and partly by reports 
of synthesis of boron nitride (BN) cages\cite{Oku00,Oku03,Stephan98,Goldberg98}.
Here,
we study  several fullerene analogues  of AlN and selected single-wall AlN nanotubes
containing up to about 200 atoms.
Some of the cage structures studied in this work have been proposed as candidate
structures for boron nitride (BN) cages synthesized and detected in mass 
spectrum\cite{Oku00,Oku03,Stephan98, Goldberg98}.
Our study, performed at the all electron density functional level,  does not 
preclude the existence of cage structures for the AlN, in agreement with conclusion 
drawn by Chang and coworkers\cite{AlN_cage2}.
  The binding energy, the 
first ionization potential, and electron affinity  are calculated.
 The
band gap calculated by difference of self-consistent solution  ($\Delta SCF $)
method show that the cages and tubes  of AlN, like their bulk phase, are 
characterized by large gap.

\section{Methodology and Computational details}
     Our calculations are performed  using the Slater-Roothaan  (SR) method\cite{Dunlap03}.
It uses Gaussian bases to fit the orbitals and effective one body Kohn-Sham potential
of density-functional theory\cite{KS65}. The SR method through robust and 
variational fitting is analytic and variational in all (orbital and fitting) 
basis sets\cite{Dunlap03}.  
The most general functionals that it can treat so far are certain 
variants\cite{Vauthier} of the X$\alpha$ functional\cite{Mess76,TJ92}.
In particular, it can handle different 
{$\alpha$'s}  on different elements analytically and variationally so that the 
atomized energies of any cluster can be recovered exactly,
and all energies are accurate through first order in any changes to any 
linear-combination-of-atomic-orbitals LCAO or fit.
The {\sl s-}type fitting bases are
those scaled from the {\sl s-}part of the orbital basis\cite{Dunlap79}. 
A package of basis sets has
been optimized\cite{GSAW92} for use with DGauss\cite{AW91}.  We use the valence double-$\zeta$ orbital basis set
DZVP2 and the {\sl pd} part of the (4,3;4,3) (A2) charge density fitting basis.
We use the $\alpha$ values of  0.748222 and 0.767473 for Al and N, respectively\cite{condmat}.
Using these $\alpha$ values 
with the above basis sets, exact total energies of Al and N atoms can be obtained.
Hence, in the present computational
model, the atoms in the cages  or tubes  will have {\em exact} atomic energies in the
dissociation limit\cite{Jhonson75}.  The accuracy of this method using the exact atomic 
values of $\alpha$ as judged from the binding energies of Pople's G2 set of molecules 
lies between that of the local density approximation 
and that of the generalized gradient approximation\cite{condmat}. Its main advantage is that 
it requires no numerical integration and hence within the accuracy of the model
gives results that are accurate to machine precision.  The optimization 
is performed using the Broyden-Fletcher-Goldfarb-Shanno (BFGS) 
algorithm\cite{BFGS1,BFGS2,BFGS3,BFGS4,BFGS5} with
forces on the atoms  computed non-recursively using the 4-j generalized Gaunt 
coefficients\cite{Dunlap02}.

\begin{figure}
\epsfig{file=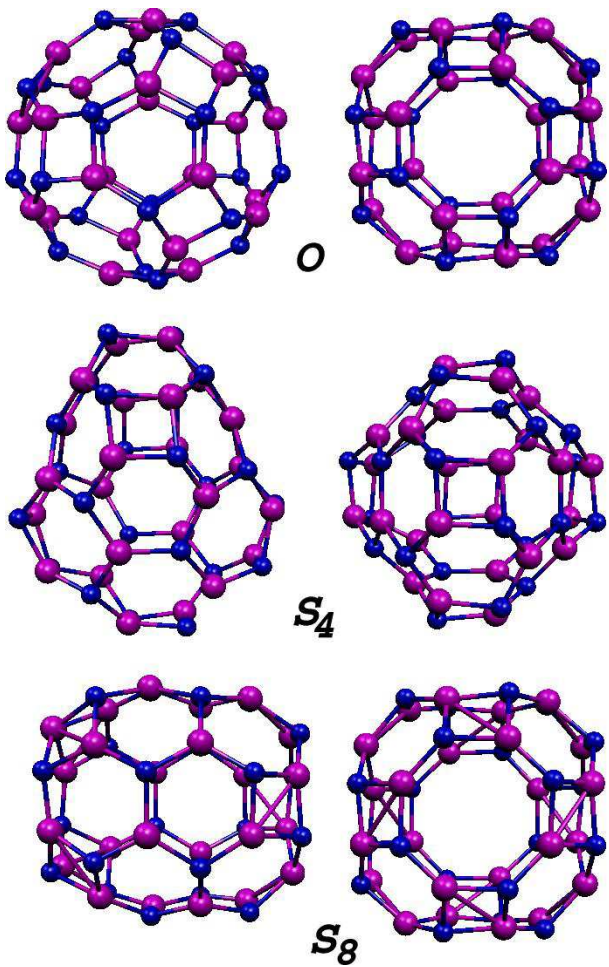,width=8.0cm,clip=true}
\caption{\label{fig:24} (Color online) Two different views of  optimized 
octahedral $O,$ $S_4,$ and \SE  \AlNF cages. }
\end{figure}

\begin{figure}
\epsfig{file=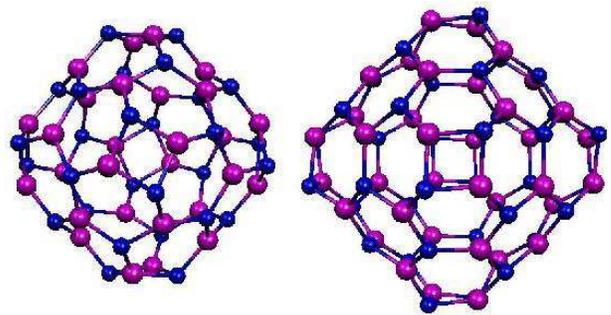,width=8.0cm,clip=true}
\caption{\label{fig:sixsq} (Color online) Favorable $T$ \AlNE and $T_d$ \AlNS cages containing six square. }
\end{figure}

\begin{table}[h!]
\caption{The bond length, dissociation energy of $^3\Pi$ AlN molecule obtained by present
analytic Slater-Roothan method in comparison with those obtained from some selected computational models and experiment. The present value of dissociation energy does not contain zero point energy.}
\label{tab:aln}
\begin{tabular}{llcl}
\hline
 Model  &   &  ~~$R_e$ (\AA)~~     &  ~~    $D_0$ (eV)~~ \\
\hline
Expt. &  Ref. \onlinecite{expt_AlN} &   1.79 &       2.86 $\pm 0.39$    \\
CCSD(T)/WMR  & Ref. \onlinecite{Gustev} &   1.79 &       2.45 \\
MRCI  & Ref. \onlinecite{Malrieu} &   1.83 &       2.42 \\
MRCI  & Ref. \onlinecite{Langhoff} &   1.81 &       2.20 \\
BP86/6-31G*  & Ref. \onlinecite{AlN_cage1} &   1.81 &       2.87 \\
BPW91/6-31G**  & Ref. \onlinecite{Pandey} &   1.80 &       2.76 \\
Present  &    & 1.81 &       2.71 \\
\hline
\end{tabular}
\end{table}

\section{Results and Discussion}

    Before we present our results on larger AlN cages, we test the computational model 
for the simple diatomic  AlN molecule, for which experimental values of bond length 
and dissociation energy are available. A number of 
theoretical calculations at different level of sophistications have also been 
performed on this molecule\cite{AlN_cage1,Gustev,Malrieu,Langhoff,Pandey}.
  The comparison of these with the numbers obtained 
in present model is given in Table \ref{tab:aln}, where the present values 
are found to be in good agreement with the experimental values. All theoretical 
values of dissociation energy are within the rather large experimental error bar.
Predicted dissociation energy in present model lies between the {\it ab initio} 
{\em coupled cluster} 
(CCSD) and {\em multi reference Configuration interaction} MRCI 
numbers on the one hand and the  Becke-Perdew86(BP86)  or Becke-Perdew-Wang91 (BPW91) 
density functional models on the other. 
Use of larger 6-311G** orbital basis and the Turbomole fitting basis 
used in our earlier work\cite{condmat,condmat1}  does not change 
the results.  Bond length is practically identical and the binding energy decreases 
by 0.2 eV. The basis set effect is largely canceled by adjusting $\alpha$ to 
get exact atomic energies.
 Thus, we expect present model to provide a
reliable description of AlN cages and nanotubes studied in this work.
 
      The optimized \AlNF clusters of octahedral, \SF and \SE symmetry are shown in Fig. \ref{fig:24}.
The octahedral cage consists of 6 octagons, 8 squares, and 8 hexagons. It is a 
round cage with two symmetry  inequivalent atoms.
This round cage has been studied as a candidate for cages made up 
of carbon\cite{DT94}, silicon hydrides\cite{LB00}, boron oxides\cite{LB00}, 
aluminum hydrides\cite{LB00} etc.  It has also been proposed as structure 
of the B$_{24}$N$_{24}$ peak, found in abundance in a recent experimental 
time-of-flight, mass spectrometric study\cite{Oku03}.
The optimized coordinates of the \AlNF cage for two inequivalent atoms 
in \AA \, are (0.868, 2.101, 3.326) for Al and (3.48, 2.13, 0.916) 
for N, respectively. Positions of other 
atoms can be determined from the symmetry group operations of an octahedral 
point group with symmetry axes along the coordinate axes.
  The cage has a radius of
4.11 \AA, about 27\% larger than that of the octahedral \BNF cage\cite{ZD04}. The Al atoms 
are on the sphere of 4.03 \AA\, radius, while the N atoms lie on the 
sphere of 4.18 \AA. These differences in radii of Al and N skeletons are of same 
magnitude to those found  for the B and N radii  in the \BNF cage.   
       In contrast, the \SF cage, also shown in Fig. \ref{fig:24}, satisfies the isolated six
square hypothesis\cite{SSL95} and does not contain any octagon. 
 The isolated square hypothesis is similar to 
the better-known isolated pentagon rule for carbon fullerenes. The exact AlN analogue of 
carbon fullerenes do not exist. Fullerenes do not permit full alternation of 
aluminum and nitrogen atoms
due to presence of pentagonal rings. Even numbered rings are necessary
in alternate cages.
Using E\"uler's theorem, it can be shown, that alternate AlN cages 
can be made closed using exactly six squares (four atom rings). The \SF structures
is such a cage containing twenty hexagons and six squares. 
It is proposed as a fullerene-like analogue of an alternate boron nitride fullerene.  

       The \SE cage contains two octagons, eight squares, and twelve hexagons.
It is indistinguishable from the \O cage  when viewed along four fold axis 
passing through the center of octagon. It is basically a very short closed 
capped $(4,4)$ armchair aluminum nitride nanotube. As it is practiced in literature
we use the term nanotube for cylindrical cage structures. Incidentally, the caps of  $(4,4)$ 
armchair AlN NT are the hemispherical halves of round octahedral \O cage\cite{ZD04}.
For boron nitride, the 
\SE and \SF cages are energetically favored over the \O cage, although the energy 
differences were quite small\cite{ZD04,ZBPD04}.  In the present \AlNF case,  similar trend 
is observed and the energy differences are further diminished.
The \SF and \SE cages are energetically nearly degenerate with \O cage 
being  higher by  0.1 eV.  

\begin{table}[h!]
\caption{The electronic structure, and symmetry of HOMO and LUMO  of 
the AlN cages and nanotubes.}
\label{tab:ES}
\begin{tabular}{lllcc}
\hline
 System    &      &  Electronic structure&   HOMO  &  LUMO\\
\hline
\AlNF    &  \sl{O}      &       10$a_{1}$ 10$a_{2}$ 30$t_{1}$ 30$t_{2}$ $20e$ & $e$ & $a_1$  \\
\AlNF    &  \sl{S$_4$}  &       $60a$ $60b$ $120e$  & $e$ & $a$   \\
\AlNF    &  \sl{S$_8$}  &       30$a$ $30b$  60$e_1$ 60$e_2$ 60$e_3$ &  $a$ & $a$ \\
\AlNE    & \sl{C$_{4h}$}&     38$a_g$ 32$a_u$ 38$b_g$ 32$b_u$ 64$e_g$ 76$e_u$  & $a_u$ & $a_g$ \\
\AlNE    & \sl{T}       &    $26a$  $44e$ $ 70t$  &  $t$  &  $a$  \\
\AlNT    & \sl {S$_8$}  &    40a 40b 80e$_{1}$ 80e$_{2}$ 80e$_{3}$   &   $b$  & $a$ \\
\AlNS    & \sl {T$_d$}  &   24$a_1$  6$a_2$ $30e$ 36$t_1$ 54$t_2$  & $t_2$& $a_1$ \\
\AlNFE    & \sl {S$_8$}  &    $60a$  $60b$ 120$e_1$ 120$e_2$ 120$e_3$  & $a$  &   $a$ \\
\AlNRA   &  \sl{O}    &     40$a_{1}$ 40$a_2$ 80$e$ 120$t_1$ 120$t_2$ & $a_2$ & $a_1$\\
\AlNRB   &  \sl{O}    &     40$a_{1}$ 40$a_2$ 80$e$ 120$t_1$ 120$t_2$  & $a_2$ & $a_1$\\
\hline
\end{tabular}
\end{table}

\begin{figure}
\epsfig{file=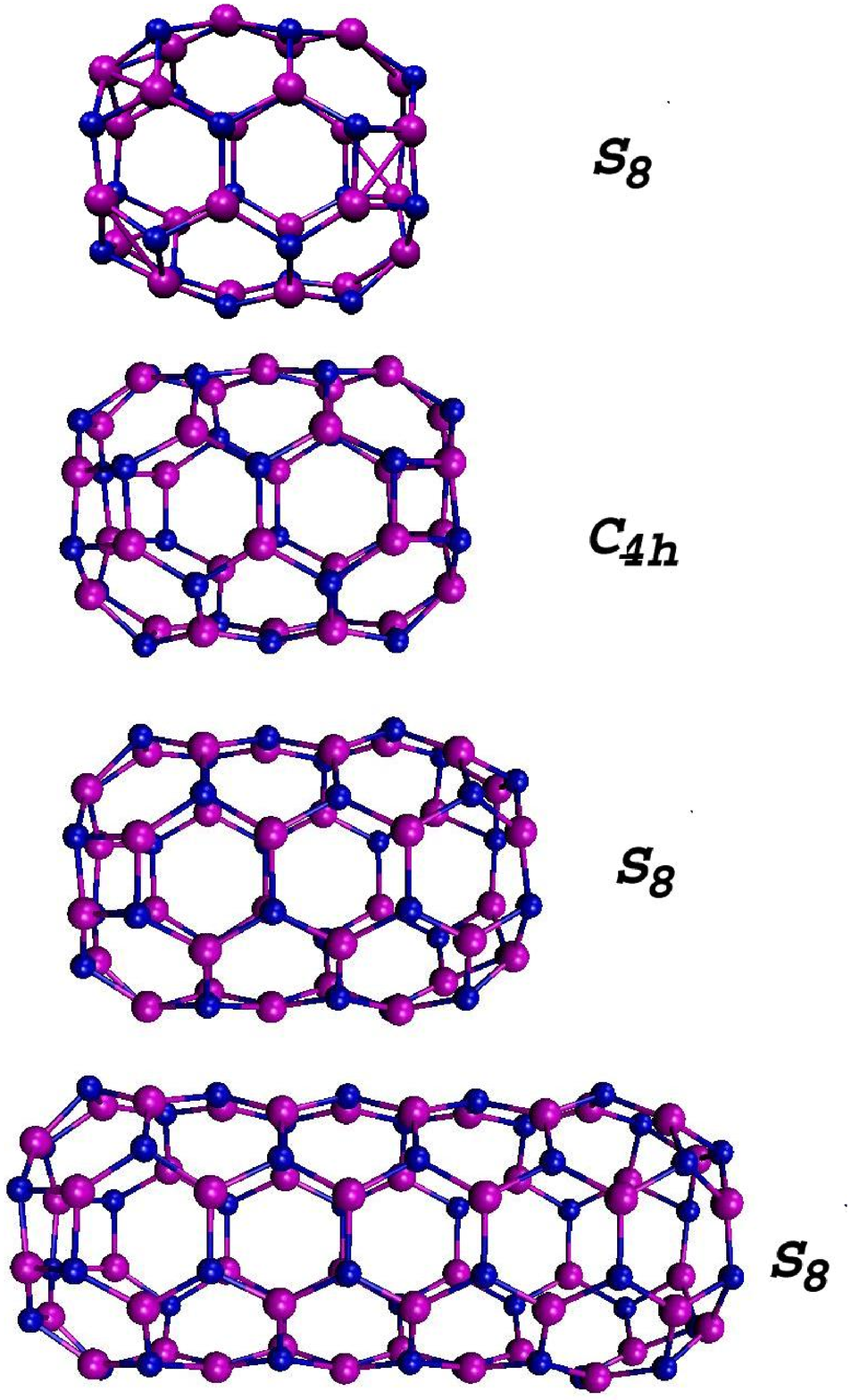,width=8.0cm,clip=true}
\caption{\label{fig:tubes} (Color online)  Finite AlN nanotubes:  \AlNF\!,  \AlNE\!,  \AlNT\!,
and \AlNFE}
\end{figure}

    The \SE \AlNF cage can be extended along the four fold axis by  inserting a 
ring of alternate AlN pairs to obtain \AlNE  nanotubes with $C_{4h}$ symmetry.
The  $C_{4h}$  \AlNE tube can also be generated from the \O \AlNF cage by cutting 
former into two halves, orienting them along the $C_4$ axis and then inserting 
a ring of eight alternate AlN atoms, followed by a rotation of one half by 
forty-five degrees.  Likewise,  the $C_{4h}$  \AlNE tube can be extended by adding 
one and five rings of alternate AlN atoms to obtain \AlNT and \AlNFE 
nanotubes, respectively. Both nanotubes have $S_8$ symmetry. 
The addition of successive alternate rings will lead to the more 
familiar infinite (4,4) AlN nanotube.  The optimized structures of the capped nanotubes 
thus derived are shown in Fig. ~\ref{fig:tubes}.   
The tubes have diameter of 6.68 \AA.  The largest tube \AlNFE 
has length of 16.5 \AA.  The AlN bond lengths in tubes in general vary from 1.74 \AA \, to 
1.80 \AA. The AlN bond that shares hexagonal and octagonal 
rings is the shortest (1.74 \AA) while the largest one (1.80 \AA) is shared by
octagonal and square rings. The inner hexagonal rings 
have AlN bond distances in the range 1.76-1.79 \AA.

      The \AlNE cage of $T$ symmetry and the tetrahedral \AlNS cage are another two 
cages that satisfy the isolated six square hypothesis, and thus prior to the 
experimental boron nitride work\cite{Oku03} were considered most favorable.  The 
former contains  twenty four
hexagons while the later has thirty-two hexagons.  The boron nitride counterpart
of both these structures have been proposed as candidate structures for 
abundant  boron nitride clusters detected in mass spectrum. The  \AlNE cage of $T$ symmetry
is energetically more favorable than the $C_{4h}$ \AlNE nanotube.
The former has six four-fold rings that isolated by hexagonal rings while the latter
has more defects eight squares and two octagons.  The extra defects in the $C_{4h}$ 
make it energetically less favorable.
Our calculations show that the AlN bond distances in the $T_d$ cage are largest (1.80 \AA) 
for the bonds that belong to the square ring. The bonds between a square and hexagon are 
1.76 \AA\, in length, and those between hexagonal rings have length of 1.79 \AA.


\begin{figure}
\epsfig{file=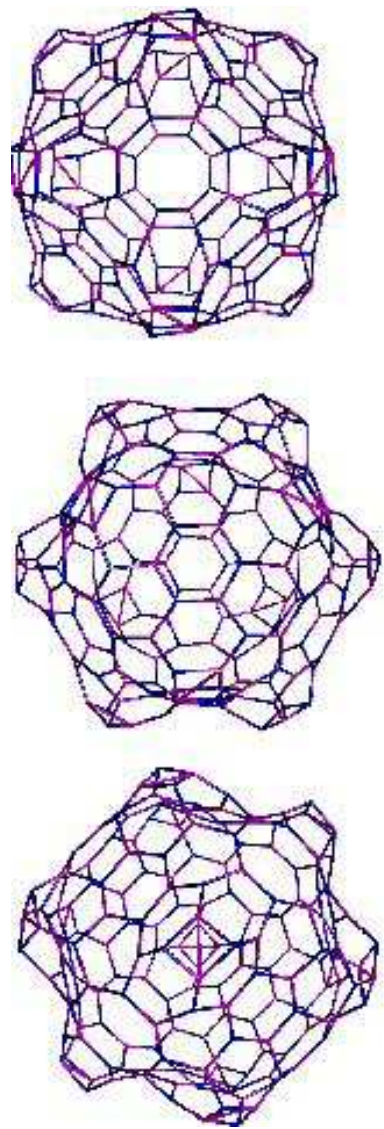,width=5.0cm,clip=true}
\caption{\label{fig:bigO1} (Color online) Octahedral  
\AlNRA cage as seen from the four-fold (top), three-fold (middle), and two-fold (bottom)
axes, respectively. }
\end{figure}

\begin{figure}
\epsfig{file=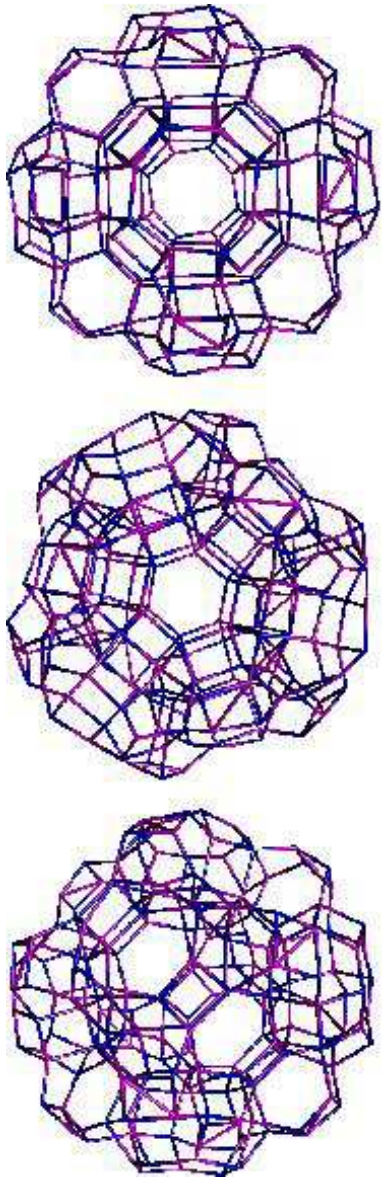,width=5.0cm,clip=true}
\caption{\label{fig:bigO2} (Color online) Octahedral  
\AlNRB cage as seen from the four-fold (top), three-fold (middle), and two-fold (bottom)
axes, respectively. }
\end{figure}

   The  \AlNR  cluster is a larger cage that preserves the symmetry of 
the \AlNF by surrounding all square and octagonal defects by rings of hexagons 
like leapfrogging in the fullerenes\cite{leapfrog}, which also leads to a cluster 
four times as big.  Leapfrog fullerenes necessarily have closed electronic shells. 
All III-V fullerene-like clusters have closed-shells due to large HOMO-LUMO gaps.
The \AlNR  cage contains eighteen defects: twelve squares and  six octagons.  
The optimization of \AlNR\!, starting from the ionic configuration of the \BNR cage 
scaled by roughly the ratio of AlN and BN average bond distances, results in the 
fullerene-like \AlNRA cage as shown in Fig. ~\ref{fig:bigO1}. 
It is evident from the figure that squares stick out. The cage has an average radius of 7.5 \AA\,
and squares protrude out by roughly 1 \AA \, from the average radius.
On the other hand, if  the optimization of \AlNR is started from the 
unscaled \BNR cage, then
one obtains a very different \O structure. The resultant structure, called 
two-shell \AlNRB cage hereafter, is shown in Fig. \ref{fig:bigO2}. It is very different from the 
\AlNRA cage in Fig. \ref{fig:bigO1} and  is basically surface growth on the 
octahedral \AlNF cage.   The \AlNF core of this structure is evident in Fig. \ref{fig:bigO2}.
The atoms on the surface of inner core have four-fold coordination.
The two-shell cage is lower in energy by 0.66 Hartree than the fullerene-like 
cage despite having many more defects than the fullerene-like \AlNRA cage.
From this result it appears that larger spherical clusters or quantum dots of 
AlN would prefer a solid-like cages  rather  than hollow fullerene like cages.
On the other hand, earlier study have found larger $Al_{60}N_{60}$ cage 
to be energetically and vibrationally stable\cite{AlN_cage2}. It would 
therefore be interesting to investigate relative stability of  
larger clusters containing fullerene-like and solid-like spherical quantum dots.
We also examined the two-shell cage as a candidate structure of \BNR. Our 
calculation indicate that the cluster is not stable and fragments upon 
optimization.

\begin{table}
\caption{The calculated values of binding energy per AlN pair (BE), the energy gap between the highest
occupied molecular orbital and the lowest unoccupied molecular orbital, the vertical ionization
potential (VIP), the electron affinity (VEA), and the energy gap obtained from the $\Delta SCF$
calculation for the optimized AlN cages. These calculations are spin-polarized.
All energies are in eV.}
\label{table:BE}
\begin{tabular}{lcccccc}
\hline
            & Symmetry  &  BE     &   GAP       & VIP      &  VEA &  $\Delta ~SCF$ \\
\hline
\AlNF        &  \sl{O}      &   10.24       &  2.97   & 7.05   &  1.46  &   5.59        \\
\AlNF       &  \sl{S$_4$}   &   10.34       &  2.47   & 6.84   &  1.72  &   5.12        \\
\AlNF       &  \sl{S$_8$}   &   10.34       &  2.63   & 6.79   &  1.58  &   5.21        \\
\AlNE        & \sl{C$_{4h}$}&   10.42       &  2.74   & 6.81   &  1.59  &   5.22        \\
\AlNE        & \sl{T}       &   10.45       &  2.67   & 6.84   &  1.69  &   5.15        \\
\AlNT        & \sl {S$_8$}  &   10.49       &  2.79   & 6.77   &  1.61  &   5.16        \\
\AlNS        & \sl {T$_d$}  &   10.54       &  2.70   & 6.73   &  1.76  &   4.95        \\
\AlNFE       & \sl {S$_d$}  &   11.09       &  2.81   & 6.56   &  1.76  &   4.8        \\
\AlNRA       & \sl {O}      &  10.56        & 2.63    & 6.22   &  2.04  &   4.12  \\
\AlNRB       & \sl {O}      &  10.75        & 1.81    & 6.03   &  2.53  &   3.50  \\
\hline
\end{tabular}
\end{table}

The electronic structures of AlN cages are given in Table ~\ref{tab:ES} while the 
binding energies per AlN pair, the HOMO-LUMO gaps, 
the ionization potential and the electron affinity  of these cages
are given in Table \ref{table:BE}. The ionization potential 
(electron affinity) is calculated from the difference in the 
self-consistent solutions of neutral AlN cage and its cation(anion).
All structures are energetically stable with binding energies of 
order of 10-11 eV per AlN pair.  It is well known that the HOMO-LUMO gaps 
obtained from the density functional models including the present 
one underestimate true band gap\cite{bandgap}.  
The so called $\Delta ~SCF$  provides good approximation for the 
ionization potential and the electron affinity of the 
system\cite{Scuseria_SIC}. We determine the band gap as
a difference of ionization potential and electron affinity computed
by the $\Delta ~SCF$ method.  The time dependent density functional theory 
(TDDFT) or the {\em GW} approximation 
provide corrections to the HOMO-LUMO gaps and are more suitable for 
calculations of the band gaps. We have not yet implemented these
techniques.  
However, the present methodology of using fitting basis sets 
can also make TDDFT calculations efficient.\cite{Rosch}
 The  $\Delta ~SCF$ calculated gaps are given in Table ~\ref{table:BE}.
In case of \AlNF although the O cage is energetically less favorable
its HOMO-LUMO gap and  $\Delta ~SCF$ gap are larger than those of the \SF and \SE 
cages.  Amongst the AlN cages that have only six isolated squares, the binding 
energy systematically increases for \SE Al$_{24}$N$_{24}$,
$T$ Al$_{28}$N$_{28}$, and  \AlNS cages.  This trend is similar to 
that observed in case of  boron nitride cages\cite{ZD04}  and carbon 
fullerenes\cite{D91}. The squares stick out too far in the 
M$_{96}$N$_{96}$ (M$=B, Al$)
cage for that cage to be favored and its hemisphere to be an extremely
favorable cap of the (8,8) nanotubes.  Both the Al$_{96}$N$_{96}$ cages 
have smaller binding energy than the   Al$_{48}$N$_{48}$ nanotube,
perhaps due to larger number of defects (octagonal and four-fold rings).
The O Al$_{24}$N$_{24}$ and Al$_{60}$N$_{60}$  cages were also studied by Change etal\cite{AlN_cage1}.
They find that former has binding energy of 4.72eV/atom while the latter has binding energy 
of 4.76 eV/atom. The increase in the binding energy 
from octahedral Al$_{24}$N$_{24}$ cage to Al$_{60}$N$_{60}$ cage is 0.05 eV/atom.
Our calculations show binding energy gain of 0.4 eV/atom with size increase 
from octahedral Al$_{24}$N$_{24}$ to octahedral Al$_{96}$N$_{96}$. It is 
therefore likely that the octahedral cages studied in this work are more stable
than the icosahedral  Al$_{60}$N$_{60}$ cage.
In comparison with their boron nitride counterparts\cite{ZD04,ZBPD04,ZBPD05}
the AlN cages are energetically less stable, have lower ionization potentials and
smaller band  gaps. This feature is similar to the bulk phase 
of these materials. The band gap of solid AlN is smaller than that
of BN solids.

   To summarize, fullerene-like cage structures and finite $(4,4)$ nanotubes 
are studied by the density functional calculations using polarized Gaussian 
basis sets of double zeta quality. The binding energy, electron affinity,
ionization potential, the HOMO-LUMO gap and the $\Delta SCF$ gap are 
calculated for the optimized AlN cages. Calculations show that all AlN cages
are energetically stable, with band gap of order of 5 eV. For the larger 
\AlNR cluster, the {\em two-shell} cage with an interior \AlNF \O cage is 
energetically favorable 
over the fullerene-like cage. The binding 
energies and band gaps are smaller than their BN counterparts. We hope that 
the present study will aid the experimental search for the AlN cages.

       The Office of Naval Research, directly and through the Naval Research Laboratory, 
and the
Department of Defense's  High Performance Computing Modernization Program, through the Common
High
Performance Computing Software Support Initiative Project MBD-5, supported this work.

\end{document}